\documentclass[12pt,aps,a4paper]{revtex4}
\usepackage{epsfig}
\usepackage{graphicx}
\usepackage{amsmath,amssymb,amsthm,color}
\usepackage[english]{babel}
\usepackage[detect-all]{siunitx}

\parskip=\medskipamount

\newcommand{\eq}[1]{(\ref{#1})}
\newcommand{\fig}[1]{Fig.~\ref{#1}}

\newcommand{\be}{\begin{equation}}
\newcommand{\ee}{\end{equation}}

\begin{document}

\title{Non-backtracking walks reveal compartments in sparse chromatin interaction networks}

\author{K. Polovnikov$^{1,2}$\footnote{To whom correspondence should be addressed. Email: kipolovnikov@gmail.com}, A. Gorsky$^{5,6}$, S. Nechaev$^{3,4}$, S. V. Razin$^{7,8}$, S. Ulyanov$^{7,8}$}

\affiliation{$^1$ Institute for Medical Engineering and Science, Massachusetts Institute of Technology, Cambridge, MA 02139 \\
$^2$ Skolkovo Institute of Science and Technology, 143026 Skolkovo, Russia \\
$^3$Interdisciplinary Scientific Center Poncelet (UMI 2615 CNRS), 119002, Moscow, Russia \\ $^4$Lebedev Physical Institute RAS, 119991, Moscow, Russia \\ $^5$ Moscow Institute for Physics and Technology, Dolgoprudnyi, Russia \\
$^6$ Institute for Information Transmission Problems of RAS, Moscow, Russia \\ $^7$ Institute of Gene Biology, Russian Academy of Sciences, Moscow, Russia \\ $^8$ Faculty of Biology, M.V. Lomonosov Moscow State University, Moscow, Russia}

\date{\today}

\begin{abstract}

%Chromatin colocalization data extracted from individual single nuclei can be rationalized in terms of sparse adjacency matrices with rich intrinsic structure.
Chromatin communities stabilized by protein machinery play essential role in gene regulation and refine global polymeric folding of the chromatin fiber. However, treatment of these communities in the framework of the classical network theory (stochastic block model, SBM) does not take into account intrinsic linear connectivity of the chromatin loci. Here we propose the ”polymer” block model, paving the way for community detection in polymer networks. On the basis of this new model we modify the non-backtracking flow operator and suggest the first protocol for annotation of compartmental domains in sparse single cell Hi-C matrices. In particular, we prove that our approach corresponds to the maximum entropy principle. The benchmark analyses demonstrates that the spectrum of the polymer non-backtracking operator resolves the true compartmental structure up to the theoretical detectability threshold, while all commonly used operators fail above it. We test various operators on real data and conclude that the sizes of the non-backtracking single cell domains are most close to the sizes of compartments from the population data. Moreover, the found domains clearly segregate in the gene density and correlate with the population compartmental mask, corroborating biological significance of our annotation of the chromatin
compartmental domains in single cells Hi-C matrices.

%Thus, the non-backtracking walks on a polymer graph is an efficient workhorse for community detection in sparse chromatin data.

\end{abstract}

\maketitle

\section{Introduction}

Many real-world stochastic networks split into self-organized communities. Social networks feature circles of friends \cite{zachary77,girvan02,newman06}, colleagues \cite{girvan02}, members of a karate club \cite{zachary77}, communities of dolphins \cite{lusseau04} etc. Cellular networks demonstrate modular organization, which optimizes crucial biological processes and relationships, such as synchronization of neurons in the connectome \cite{harris03,humphries11}, efficiency of metabolic pathways \cite{jeong00,ravasz02}, genes specialization \cite{zhang05} or interaction between enhancers and promoters \cite{doyle14}.

Interest to polymer modular networks has appeared recently in the context of genome spatial folding. Proximity of chromatin loci in space is believed to be deeply connected with gene regulation and function. Hi-C experiments \cite{dekker02,lieberman09,nagano12} provide the genome-wide colocalization data of chromatin loci. As the main outcome of the experiment, large genome-wide matrices of contacts from each individual cell or from the population are produced. Analyses of these matrices has revealed that the eukaryotic genome is organized in various and biologically relevant communities, whose main function is to insulate some regions of DNA and to provide easy access to the others. In particular, the data collected from a population of cells suggest that transcribed ("active") chromatin segregates from the, "inactive" one, forming two compartments in the bulk of the nucleus \cite{lieberman09,fortin15}. Within compartments chromatin is organized further as a set of topologically-associated domains (TADs) \cite{dixon12,sexton12, szabo19} that regulate chromatin folding at finer scales. However, interpretation and validation of communities in individual cells remains vaguely defined due to sparsity of respective data.

The broad field of applications of stochastic modular networks has initiated the boost development of community detection methods. Spectral algorithms exploit the spectrum of various operators (adjacency, Laplacian, modularity) defined on a network to identify the number of communities and to infer the optimal network partition \cite{fortunato10,newman13,newman03,shen10,decelle13}. Typically, leading eigenvectors of these operators positively correlate with the true community structure or with the underlying core-periphery organization of the network \cite{polovnikov_btc}. These algorithms, along with the majority of theoretical results in the field, are derived for the stochastic block model (SBM) \cite{fortunato10,decelle13} as an extension of  Erd\"{o}s-R\'{e}nyi graphs \cite{erdos59} to graphs with explicitly defined communities. One of the strongest limitations of the SBM is that edges between vertices belonging to the same cluster inevitably attain equal weights. At the same time, biological networks typically have several levels of organization within their communities \cite{ravasz03}. In particular, identification of several hierarchical levels in the network becomes tremendously important in the case of polymer networks, where different pairs of loci have marginally different probabilities to form a contact in space \cite{lee19}, caused by the frozen linear connectivity along the chain.

Even for simplest polymer systems the contact probability demonstrates a power-law behavior with the dimensional-dependent scaling exponent characterizing universal long-ranged behavior of polymer folding \cite{gros_khok}. In this work we propose the "polymer stochastic block model" which reflects a specific global polymer network organization with explicit structuring into communities. The main new ingredient of the model under consideration is the average contact probability $P(s=|i-j|)$ between the pairs of loci $(i,j)$ which is constant for standard non-polymeric networks, however cannot be neglected for polymers. %We derive the generalized modularity functional for the model and establish its connection with the maximum entropy principle.

Chromatin single cell networks are not only polymeric, but also sparse \cite{nagano12,flyamer17}. It is known that upon reduction of the total number of edges in the network, there is a fundamental resolution limit for all community detection methods \cite{decelle13,zhang12}. Furthermore, traditional operators (adjacency, Laplacian, modularity) fail far above this resolution limit, i.e. their leading eigenvectors become uncorrelated with the true community structure above the threshold \cite{krzakala13}. That is explained by emergence of tree-like subgraphs (hubs) overlapping with true clusters in the isolated part of the spectrum for these operators. The edge of the spectral density of sparse networks is universal and demonstrates the so-called "Lifshitz tail" \cite{polovnikov_ufn,goh01,nadakuditi13,lif1}. Localization on hubs, but not on true communities is a drawback of all conventional spectral methods in the sparse regime.

To prevent the effect of localization on hubs and to make spectral methods useful in sparse regime, Krzakala et al. proposed to deal with non-backtracking random walks on a directed graph that cannot revisit the same node on the subsequent step \cite{krzakala13}. The crucial property of  non-backtracking walks \cite{hashimoto89} is that they do not concentrate on hubs. It has been shown that the non-backtracking operator is able to resolve the community structure in sparse stochastic block model up to the theoretical resolution limit. Typically, the majority of eigenvalues of the non-backtracking operator (which is a non-symmetric matrix with complex eigenvalues) are located inside a disc in a complex plane, and a number of isolated eigenvalues lie on the real axis. %Recently M. Newman has proposed a non-backtracking flow operator which conserves probability at each node \cite{newman13}. Other versions have been recently proposed \cite{singh14} to tackle shortcomings of the non-backtracking (e.g. ignorance of hanging trees).

For the sake of community detection in sparse polymer networks we construct the polymer-type non-backtracking walks, appropriate for community detection in graphs with hidden linear memory ("polymeric background"). We establish the connection between this operator and the generalized polymer modularity, thus, bridging a gap with the maximum entropy principle. We test the performance of different spectral methods (with and without polymer background) on sparse artificial benchmarks of polymer networks that mimic compartmentalization in single cell Hi-C graphs. We show that polymer non-backtracking walks resolve the structure of communities up to the detectability threshold, while all other operators fail above it. In order to demonstrate efficiency of the method on real data, we partition a set of single cell Hi-C contact maps of mouse oocytes into active (A) and inactive (B) compartments by different operators. Found domains are shown to have similar sizes to the compartmental domains and correlate with the compartmental mask from the population-averaged data. Analyses of the GC content within the domains demonstrates enrichment and depression of the genes density in the two clusters, thus, corroborating their biological significance.

The structure of the paper is as follows. In Section II we propose the polymer stochastic block model, derive the entropy and the corresponding generalized modularity functional. In Section III we discuss polymer non-backtracking walks, prove their robustness on the benchmarks emulating compartments, and, finally, test them on the real single cell data. In Section IV we draw the conclusions.

\section{Stochastic block model with polymer contact probability}
\subsection{Definition of the model}

Characterize a $N$-bead polymer chain by coordinates $\{x_1,x_2,...,x_N\}$ of monomers $i = 1,2, ..., N$ and construct a corresponding topological graph ${\cal G}=(V, E)$ with the adjacency matrix $A_{ij}$ (accounting for the bead's proximity in space). Such graphs are typically constructed upon processing of chromatin single cell Hi-C data and in computer simulations of DNA folding \cite{dekker02,lieberman09}. A graph ${\cal G}$ does not contain pairwise spatial distances of the polymer configuration, however, provides information on spatial proximity of monomers (or groups of monomers), which is usually of major biological relevance. For the 1-bin resolution of ${\cal G}$ the polymer beads (bins) are the nodes $V$. The edge between a pair of nodes $(i, j)$ is defined by the condition $(i,j) \in E$ iff $|x_i-x_j| < \varepsilon$, where the threshold $\varepsilon$ is some cutoff radius with which the contacts between the two loci are registered in Hi-C. Due to finite excluded volume of chromatin, the theoretical number of contacts per monomer that can be registered in single cell experiments is of order of few units, while the total size of the polymer chain, measured in number of beads, is huge ($N \sim 10^5$ in the 1-kb resolution for human chromosomes). Thus, the single cell contact matrices are essentially sparse \cite{nagano12, flyamer17}. Summation over realizations of adjacency matrices $A_{ij}$ obtained from different cells results in a "population-averaged" matrix ${\cal A}_{ij}$. By construction, entries of the weight matrix ${\cal A}_{ij}$ are proportional to the probability that the spatial distance between monomers $(i,j)$ is less than $\varepsilon$.

Already for the simplest configurations, such as a conformation of ideal polymer chain isomorphic to the random walk, the matrix ${\cal A}_{ij}$ is not expected to be uniform. This is due to a polymeric power-law behaviour of a contact probability,
\be
P(s) \sim s^{-\alpha}, \quad \mbox{for $s=|i-j|$}
\label{ps}
\ee
By definition, $P(s)$ is probability to find two beads of a linear chain, separated by a chemical distance $s$, close to each other in space. The critical exponent, $\alpha$, is an important parameter, which characterizes the "memory" about the embedding of a polymer loop of length $s$ in a $D$-dimensional space \cite{gros_khok}. Such a memory can arise due to some equilibrium topological state of chromatin, or could be a result of partial relaxation of mitotic chromosomes \cite{rosa-everaers}. Notable examples of $\alpha$, typically appearing in the chromatin context for chain embedding in a three-dimensional space, are $\alpha=3/2$ for ideal chain and $\alpha\approx 1$ for the crumpled globule \cite{grosberg93,grosberg88,polovnikov, lieberman09}.

Communities of folded chromatin refine the background (polymeric) contact probability at small scales and are biologically significant.
%The inactive chromatin with low transcriptional activity get clustered, giving rise to global segregation of chromatin in two compartments in the nucleus \cite{lieberman09,fortin15}. Their signature is seen as a checkerboard pattern on the population-averaged Hi-C contact maps. Apart of compartments, there are other types of smaller clusters featuring chromatin folding (topologically-associated domains \cite{dixon12, sexton12,szabo19}, polycomb domains \cite{eagen17} etc.) and their mutual relationship and cross-regulation yet is not fully understood.
We treat communities as canonical stochastic blocks \cite{fortunato10,decelle13} superimposed over the background. Stochastic block model is a network model in which $N$ nodes of a network are split into $q$ different groups $G_i$, $i = 1,2, ..., q$ and the edges between each pair of nodes are distributed independently with a probability, depending on the group labels ("colors") of respective nodes. In a matrix of pairwise group probabilities $\Omega = \{\omega_{rt}\}$ with $(r,t) = 1, 2, ..., q$, any randomly chosen pair of nodes $(i,j)$ (where $i \in G_r, j \in G_t$)  is linked by an edge with probability $\omega_{rt}$. The corresponding entry in the adjacency matrix $A_{ij}$ is $1$ with probability $\omega_{rt}$ and $0$ otherwise. The sum of many such "single-cell" Bernoulli matrices generates an analogue of the "population-averaged" Hi-C matrix ${\cal A}_{ij}$ with Poisson distributed number of contacts with the mean $\langle {\cal A}_{ij} \rangle = \omega_{rt}$ where $i \in G_r, j \in G_t$. To the first approximation, the communities can be considered identical (known as a "planted" version of the model)
\be
\Omega_{rt} = \Bigg\{\begin{array}{lr} w_{in}, & r=t \\ w_{out}, & r \ne t \end{array}
\label{omega}
\ee

Having \eq{ps} and \eq{omega}, the simplest assumption one can come up with is that formation of compartments in chromatin is independent of the global memory of folding. Indeed, phenomenon of compartments is likely related to preferential interactions of nodes of the same epigenetic type (e.g., "active" or "inactive") and is modelled as a phase separation of block-copolymers \cite{nuebler}. This allows to suggest the factorization of \eq{ps} and \eq{omega}, so that the final probability for the edge $(i,j)$ reads
\be
Prob_{ij} = P(|i-j|) \Bigg\{\begin{array}{lr} \omega_{in}, & r = t \\ \omega_{out}, & r \ne t
\end{array}, \quad i \in G_r,\; j \in G_t
\label{prob}
\ee

\begin{figure}[ht]
\centerline{\includegraphics[width=14cm]{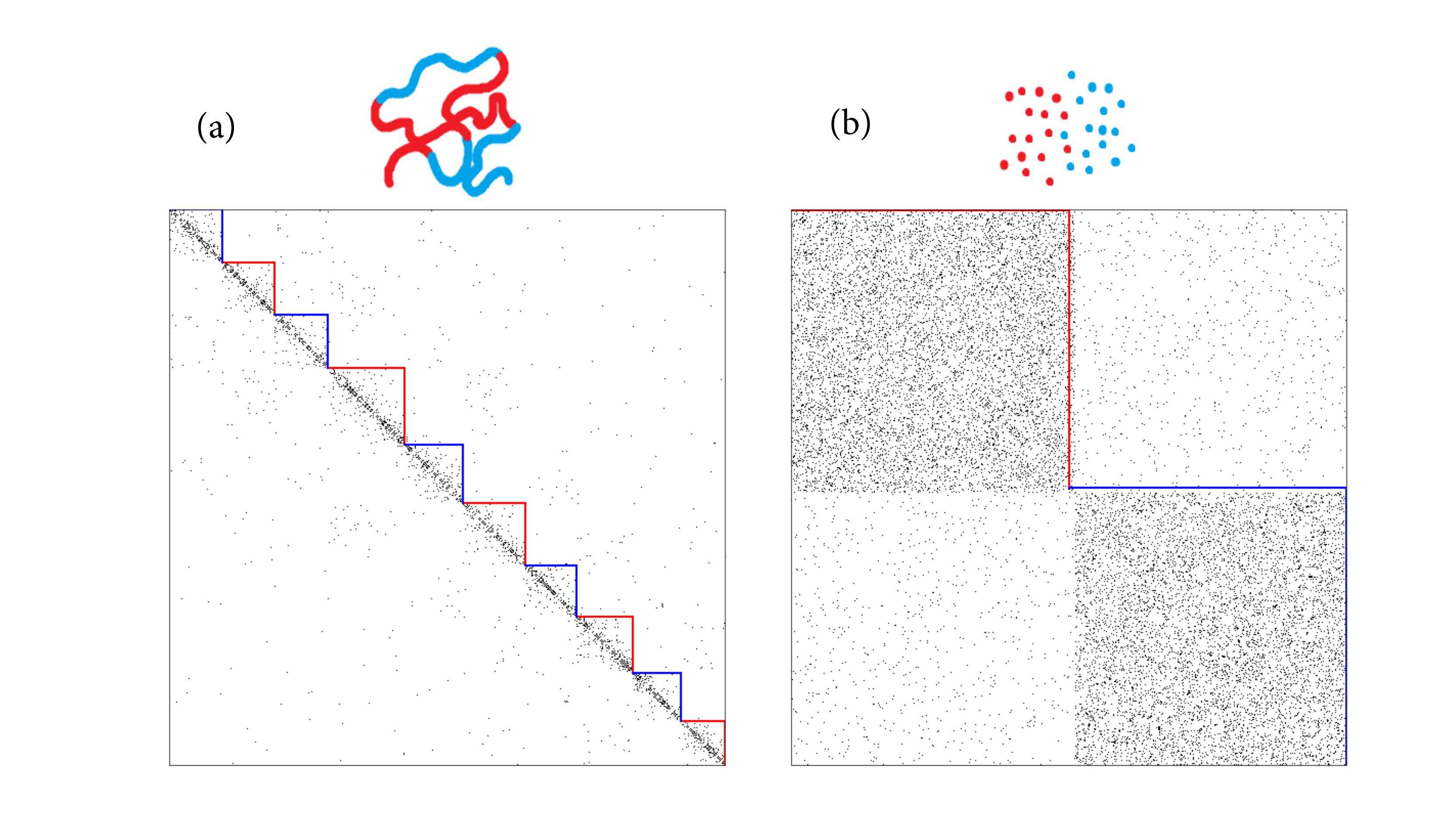}}
\caption{Adjacency matrices of $N=1000$ with \textit{two clusters} generated according to the (a) polymer stochastic block model ($w_{in}=1, w_{out}=0.1, P(s)=s^{-1}, \lambda=100$) and (b) canonical stochastic block model ($w_{in}=0.1, w_{out}=0.01, \lambda=500$). Vertices in the graph are enumerated by the polymer coordinate (a) and first all red, then all blue ones (b).}
\label{fig:01}
\end{figure}

To emulate A and B compartments in a single cell Hi-C network, we consider a simple adjacency benchmark of a polymer with two communities. Namely, we represent the chain as a sequence of alternating segments of $A$ and $B$ type (painted in red and blue), whose lengths are Poisson-distributed with the mean length $\lambda$. An example of the resulting adjacency matrix is depicted in \fig{fig:01}(a). Note that due to decay of the contact probability, the "checkerboard" compartmentalization pattern is hardly seen in single cells Hi-C data \cite{flyamer17}. Since segments of the same type are surrounded in space by segments of the other type, they form local "blob-like" clusters along the main diagonal of the adjacency matrix reminiscent to topologically-associated domains \cite{dixon12}. However, they are likely formed by a different mechanism and have an order of magnitude larger size than TADs \cite{nuebler}. Such a multi-domain blob structure in \fig{fig:01}(a) is a manifestation of the polymeric nature of the network and it cannot be reproduced with communities of general memory-less networks, i.e. in the framework of the canonical stochastic block model with two clusters -- see \fig{fig:01}(b) for comparison.

\subsection{Statistical inference of polymer SBM and generalized modularity functional}

%Here we perform the statistical inference of communities assuming that the background probability is known.
Suppose that a population-averaged matrix ${\cal A}$ is observed. By definition, each entry ${\cal A}_{ij}$ of this matrix counts the amount of reads between the bins $i$ and $j$ coming from a population of single cells. Thus, after proper normalization, $A_{ij}$ is a Poisson variable with the mean dictated by \eq{prob}, $\langle A_{ij} \rangle=P_{ij} \;\omega_{g_i g_j}$, and $\omega_{g_i g_j} = \Omega_{ij}$ are the pairwise group probabilities (at the moment we do not require all the groups to be identical). Neglecting correlations between the matrix entries, the statistical weight of ${\cal A}$ conditioned on the cluster probability matrix $\Omega$, background contact probability $P$ and group labels of the nodes $\{g_i\}$, can be factorized into the product of the Poisson probabilities for the entries $A_{ij}$
\be
Z({\cal A}|\; \Omega, P, \{g_i\}) = \prod_{i<j} \frac{\left(P_{ij}\; \omega_{g_i g_j}\right)^{{\cal A}_{ij}}}{{\cal A}_{ij}!} \exp\left(-P_{ij}\; \omega_{g_i g_j}\right)
\label{p}
\ee
where the product runs over all pairs of nodes in the network. Since there are no self-edges in the network, all the diagonal elements of the matrix ${\cal A}_{ij}$ are zeros and we do not include them into the product \eq{p}. The corresponding partitioning entropy of the polymer SBM is
\be
\log Z({\cal A}|\; \Omega, P, \{g_i\}) = \sum_{i<j} \left({\cal A}_{ij} \log \omega_{g_i g_j} - P_{ij}\;\omega_{g_i g_j}\right)
\label{entr}
\ee
where we have omitted the constant terms $-\log {\cal A}_{ij}!$ and ${\cal A}_{ij} \log P_{ij}$, independent of the partitioning. For identical communities (see \eq{omega}), we get
\be
\begin{cases}
\omega_{g_i g_j} = w_{out} + \delta_{g_i g_j}\left(w_{in} - w_{out}\right) \\
\log \omega_{g_i g_j} = \log w_{out} + \delta_{g_i g_j}\left(\log w_{in} - \log w_{out}\right)
\end{cases}
\label{planted}
\ee
Taking \eq{planted} into the account and omitting again all irrelevant constant terms, we arrive at the final expression for the entropy \eq{entr}
\be
T \log Z({\cal A}|\;\Omega, P, \{g_i\}) = \sum_{i<j} \left({\cal A}_{ij} - \gamma P_{ij}\right)\delta_{g_i g_j}
\label{logp}
\ee
where $T = \left(\log w_{in} - \log w_{out}\right)^{-1}$ is the effective temperature and
\be
\gamma = \frac{w_{in} - w_{out}}{\log w_{in} - \log w_{out}}
\label{gamma}
\ee
is a parameter describing the cluster probabilities inherited from the initial definition of  stochastic blocks. %The same resolution parameter has appeared previously for the degree-corrected planted stochastic block model with $P_{ij} = d_i d_j/2m$, where $d_i$ is a degree of the $i$-th node and $2m = \sum_{i} d_i$ is twice the total number edges in the network \cite{newman16}.

The entropic functional \eq{logp}, up to normalization coefficients and constant terms, is the generalized modularity functional. For $P_{ij} = d_i d_j/\sum_i d_i$, where $d$ is the vector of degrees, \eq{logp} reduces to the modularity proposed by M. Newman \cite{newman16,newman06} for the sake of spectral community detection in scale-free networks. Recently it has been shown that the same functional can be used to partition a network with the core-periphery organization \cite{polovnikov_btc}. The operator of the generalized modularity reads
\be
\mathbf{Q} = A - \gamma P
\label{q1}
\ee
The second term in \eq{q1} can be understood as an expectation number of contacts between nodes $(i,j)$ in the population-averaged data, or as a probability of the link in the single cell graph. Indeed, in absence of the stochastic blocks, this value equals $P_{ij}$ by definition. The factor $\gamma$ responds for the clustering structure superimposed over the background. In the limit of "weak" communities, when $w_{in} = w_{out} \to 1$, the partitioning yields $\gamma \to 1$, which corresponds to the pure background. To determine the optimal value of $\gamma$, one can run a recursive procedure, which consists of iterative maximization of the generalized modularity and renormalization of $\gamma$ according to \eq{gamma}. We realize this approach in our numerical analyses below.

\section{Polymer non-backtracking flow operator}

\subsection{Non-backtracking walks on a directed polymer network}

Search for the global maximum to the modularity functional is a very hard computational problem. One of most promising approaches which avoids a brute force, is to suggest that if the community structure is significantly strong, there is an operator whose eigenvectors encode the network partitioning in these communities \cite{newman06,decelle13}. However, as it was first noted by Krzakala et al \cite{krzakala13}, for sparse networks leading eigenvectors become uncorrelated with true community structure well above the theoretical threshold. As a result, all conventional operators such as adjacency, Laplacian and modularity fail to find communities in rather sparse networks.

To overcome this difficulty, it was proposed to exploit the spectrum of the Hashimoto matrix $\mathbf{B}$, which is a transfer matrix of non-backtracking walks on a graph \cite{hashimoto89}. It is defined on the edges of the directed graph, $i\to j, k\to l$, as follows
\be
\mathbf{B}_{i \to j, k\to l} = \delta_{il}(1-\delta_{jk})
\label{b}
\ee
It is seen from \eq{b} that the non-backtracking operator prohibits returns to the point which a walker has visited at the previous step. Since matrix $\mathbf{B}$ is non-symmetric, its spectrum is complex. For Poissonian graphs the spectral density of $\mathbf{B}$ is constrained within a circle of radius $\sqrt{\langle d \rangle}$ in the complex plain and exhibits no "Lifshits tail" singularities near the spectral edge, in contrast to other conventional operators \cite{polovnikov_ufn,krzakala13}. Real eigenvalues of $\mathbf{B}$ lying out of the circle become relevant to the community structure even in sparse networks. Associating the corresponding eigenvectors with the network partitioning permits to detect communities all the way down to the theoretical limit. In \cite{newman13} M. Newman suggested a normalized operator, that conserves the probability flow at each step of the walker.

%In \cite{newman13} M. Newman suggested to use a non-backtracking flow operator, which normalizes probability at each node.
For the sake of community detection in sparse polymer graphs, we propose a conceptually similar operator that describes the evolution of the non-backtracking probability flow on a graph with intrinsic linear memory
\be
\mathbf{R}_{i \to j, k \to l} = \frac{\delta_{il}\left(1 - \delta_{jk}\right)}{d_i - 1} - \gamma \left(d_j d_l \right)^{-1} P_{jl}
\label{flow}
\ee
In Appendix we establish the connection between the non-backtracking operator and the generalized modularity, derived in the previous Section from the statistical inference of the polymer SBM. Thus, partitioning of a polymer network into two communities according to the leading eigenvector of the polymer non-backtracking flow operator \eq{flow} responds to the maximum entropy principle.

\begin{figure}[ht]
\centerline{\includegraphics[width=14cm]{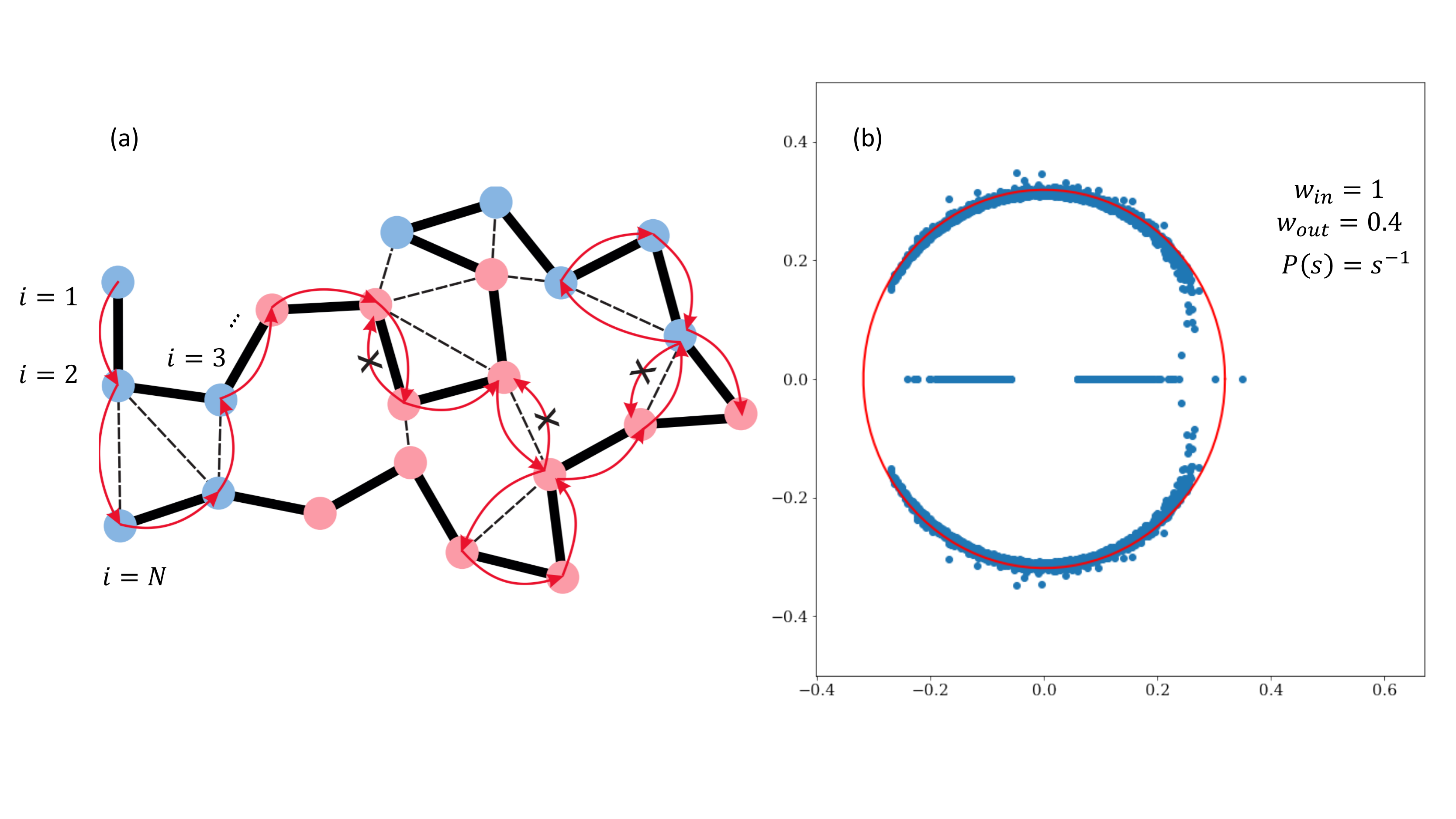}}
\caption{(a) Depiction of the polymer SBM network: the backbone (bold), contacts between genomically distant monomers (dashed) and two chemical sorts of the monomers (red and blue), arranged into contiguous alternating segments. An example of the non-backtracking walk on such graph is shown by arrows. Immediate returns are forbidden, preventing localization on hubs; (b) Spectrum of the polymer non-backtracking flow \eq{flow} for the fractal globular ($P(s)=s^{-1}$) large-scale organization of the chain with two overlaid compartments with the mean length $\lambda=100$.}
\label{fig:02}
\end{figure}

An example of the non-backtracking walk on a polymer graph is illustrated in the \fig{fig:02}(a). Note that despite immediate revisiting of nodes is forbidden, the walker is allowed to make loops. The second term in \eq{flow} plays a role of neutralization towards the contact probability, arising from the linear organization of the network. This compensation provides a measure for the non-backtracking operator to tell apart the true communities from the fluctuations, evoked by the polymeric scaling. Trivially, the proposed non-backtracking operator is converged to the Newman's flow operator, when the background is non-polymeric, but rather corresponds to the configuration model with fixed degrees $H_{ij} = d_i d_j/2m$ \cite{newman13}. For a pure polymeric graph without contamination by communities, the spectrum of \eq{flow} lies inside a circle of radius $r = \sqrt{\langle d (d-1)^{-1}\rangle}$. As sufficiently resolved communities are formed in the network, isolated eigenvalues pop up at the real axis.

In \fig{fig:02}(b) we depict the non-backtracking spectrum of a polymer SBM, corresponding to the fractal globule polymer network with $P(s)=s^{-1}$ of the size $N=1000$ with two compartments, organized as contiguous alternating segments with the mean length $\lambda=100$. For the parameters $w_{in}$, $w_{out}$ used, the two compartments are well resolved that is provided by the isolated eigenvalue separated from the circle. Since the leading eigenvector $u^{(1)}$ of the polymer non-backtracking flow, in contrast to the adjacency or modularity, is defined on directed edges of the network, one needs to evaluate the Potts spin variables $g_i=\pm 1$ in order to classify the nodes. From the correspondence between the modularity and polymer flow operator one sees that contribution to the $i$-th node $g_i$ comes from the flow along all the directed edges pointing to $i$. Thus, in order to switch from edges to nodes, one needs to evaluate the sign of the sum $v_i = \sum_j A_{ij} u_{j\to i}^{(1)}$ and to assign the node $i$ accordingly, $g_i=sign(v_i)$.

\subsection{Spectral clustering of the polymer stochastic block model}

In this section we investigate spectral properties of the polymer non-backtracking flow and compare performance of various linear operators in partition the polymer SBM. The two compartments with $\lambda=100$ are superimposed over the fractal globule, $P(s)=s^{-1}$, with total size of the network, $N=1000$. We fix the weight of internal edges at $w_{in} = 1$ and change the resolution of compartments by tuning the weight of external edges, $w_{out} = \numrange[range-phrase = -]{0.1}{0.8}$. Efficiency of splitting is assessed by the fraction of correctly classified nodes.

\begin{figure}[ht]
\centerline{\includegraphics[width=16cm]{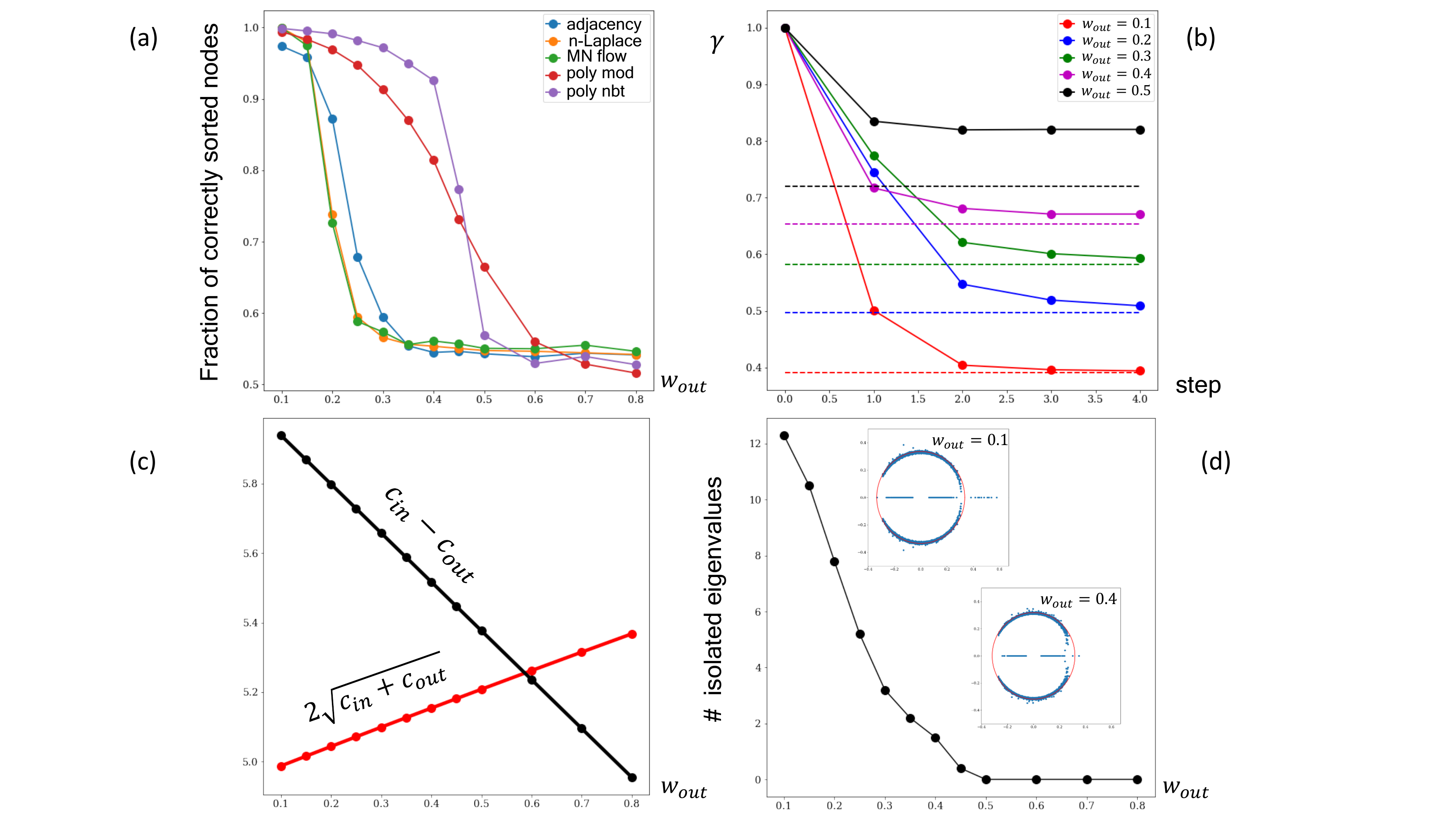}}
\caption{(a) Comparison of performance of different classical operators without background, polymer modularity and polymer non-backtracking flow operators ($N=1000, P(s)=s^{-1}, w_{in}=1, \lambda=100$); (b) The iterative approach that can be used to determine the optimal value of $\gamma$ for five values of $w_{out}$; the true optimal values of $\gamma$ calculated from \eq{gamma} are shown by dash; (c) The mean numbers of inner $c_{in}$ and outer $c_{out}$ edges are calculated for each value of $w_{out}$ in order to estimate the detectability threshold for the corresponding regular network. (d) Amount of isolated eigenvalues of the polymer flow operator plotted against $w_{out}$. Full spectra of the polymer flow operator for the two values of $w_{out}$ are shown in the insets.}
\label{fig:03}
\end{figure}

In \fig{fig:03}(a) we compare the performance of adjacency, normalized Laplacian, M. Newman's non-backtracking flow operator, polymer modularity and polymer non-backtracking flow matrices. For the latter two, the optimal value \eq{gamma} of the parameter $\gamma$ was chosen. It is evident that the polymer flow operator surpasses all conventional operators without the background, as well as the polymer modularity everywhere below $w_{out}\approx 0.5$. Qualitatively similar behaviour was demonstrated by the traditional non-backtracking operator without the background, when it was compared to other operators in \cite{krzakala13}. Therefore, our analyses (i) underscores the importance of taking into account the contact probability (polymer background) when dealing with polymer graphs, and (ii) recapitulates efficiency of non-backtracking walks in resolving communities in sparse networks.

It is worth noting that the abrupt fall in performance of the polymer flow operator coincides with the leveling of its amount of isolated eigenvalues at zero, see \fig{fig:03}(d). Values around $w_{out}\approx 0.5$ define the detectability transition, above which the leading eigenvector becomes uncorrelated with the true nodes assignment. To understand whether it corresponds to the theoretical detectability limit, we translate $w_{out}$ into the average amount of inner, $c_{in} = N w_{in}/2$, and outer, $c_{out} = N w_{out}/2$, edges and plot them as functions of $w_{out}$. As it is shown in \fig{fig:03}(c)), the polymer flow operator drops close to the theoretical detectability transition for regular stochastic block models \cite{radicchi13} (i.e. each node has \textit{exactly} $c_{in}$ random links with other nodes in its community and \textit{exactly} $c_{out}$ randomly pointed links to nodes from the other community)
\be
c_{in}-c_{out} > 2\sqrt{c_{in}+c_{out}}
\label{det}
\ee

For the stochastic block model the number of isolated eigenvalues of $\mathbf{B}$ exceeds the number of communities by one \cite{krzakala13}. However, in case of the polymer operator $\mathbf{R}$ the number of isolated eigenvalues can be much larger and "apparent" clusters might be formed "locally" at the main diagonal due to the frozen linear connectivity, see \fig{fig:01}(a). This is evident from the \fig{fig:03}(d), which shows that the number of isolated eigenvalues for the polymer flow operator can be of order of the amount of the segments ($N/\lambda$), if $w_{out}$ is sufficiently low. Indeed, for the fractal globule probability of the edge between two distant segments of the same type is $s$ times smaller than probability of the link for two close monomers ($s=|k-m|$ is the genomic distance between segments $k$ and $m$). Due to the overall small number of contacts in the network, the polymer non-backtracking flow ends up rationalizing them as separate clusters.

The value of $\gamma$ cannot be chosen arbitrary since it characterizes optimal parameters of  stochastic blocks. Thus, one may propose the following iterative approach:
\begin{itemize}
\item[(i)] begin with the initial value $\gamma_0=1$, for which we obtain the network partition;
\item[(ii)] use the amount of inner and outer edges for estimating $w_{in}$, $w_{out}$;
\item[(iii)] recalculate  $\gamma_1$ according to \eq{gamma};
\item[(iv)] repeat the procedure iteratively until $\gamma$ converges to $\gamma_{opt}$.
\end{itemize}
Results of this procedure are demonstrated in the \fig{fig:03}(b) for five different values of $w_{out}$. It is seen that just several steps of iteration is sufficient to obtain a reasonable convergence towards the theoretical values provided by \eq{gamma}. A drawback of this iterative procedure is that at each step one needs to evaluate the spectrum of the operator $2m\times 2m$, which could become a hard computational task for large and dense networks. As a reasonable approximation to the optimal value of $\gamma$ for the polymer flow operator, one can evaluate $\gamma_{opt}$ similarly for the polymer modularity, which is smaller in size and symmetric.

%An alternative method based on the analyses of the contact probability $P(s)$ of the polymer network can be suggested. At small scales $s$ the major contribution to the contact probability comes from contacts within the segments of same type (A or B), therefore, $P(s)\approx w_{in} s^{-1}$. On the other hand, when the separation distance between the loci increases up to the segment average size, $\lambda$, contacts between distinct clusters prevail and one has $P(s)\approx w_{out} s^{-1}$, see the inset in \fig{fig:03}(b). Using these simple arguments, one can deduce values of $w_{in}, w_{out}$ and $\gamma$ from the contact probability. Though, this method might become imprecise in case of non-Poissonian networks or in networks with broad distribustion of compartments size.

\subsection{Polymer non-backtracking flow resolves compartments in a single cell Hi-C network}

To check robustness of the polymer non-backtracking flow operator on real Hi-C data we run it on a set of individual oocyte cells of mouse \cite{flyamer17}. From the public repository we have taken the single cells Hi-C data on cis-contacts of 20 chromosomes from 13 single cells (260 adjacency matrices, in total). While single cells matrices with sufficiently large number of contacts are not sparse and can be split into compartments using conventional methods largely used for the bulk data (e.g., the leading eigenvectors of observed/expected transformation of a population-averaged Hi-C map, \cite{lieberman09}), here we take the cells with low to moderate amount of contacts for the sake of comparative analyses of clustering performance of different spectral methods on \textit{sparse} polymer graphs.

Before proceeding with the analyses of compartments in single cells, the raw data must be preliminary processed. In order to extract compartmentalization signal from the maps, we have coarse-grained them to the resolution 200kb. At this resolution all finer genome folding structuring  (like topologically-associated domains) is encoded within the coarse-grained blobs and does not communicate with two large-scale $A$ and $B$ compartments. We note, that, in principle, the method is applicable at higher resolution as well. However, there are two important considerations. The non-backtracking operator is defined on the edges, therefore, the leading eigenvectors need to be computed for much larger matrix than in case of traditional operators, which are defined on the nodes (e.g., modularity). This means that the computation time of the method is very sensitive to the resolution. Furthermore, one needs to be very careful with the overall network density: it decreases by several times upon decreasing of the bin size, so that one can occasionally cross the detectability limit \eq{det}. In each particular case the resolution for the annotation should be chosen with respect to the sparsity of the experimental single cell contact maps. According to this logic, we have decided to use the resolution 200 kb for the data of Flyamer et al.

Most of the contacts in the cells have degeneracy 1 at the chosen resolution, however, several pairs of bins have more than 1 contact. To preserve this feature of enhanced connectivity, we consider the counts of contacts between the pairs as weights of the corresponding edges. Furthermore, the single-cell maps are noisy and some of really existing contacts get lost due to technical shortcomings of the experimental protocol. As long as the neighboring blobs in the chromatin chain are connected with probability 1, all lost contacts $A_{i,i+1}$ need be added to the adjacency matrix manually; we assign the weight 1 to such edges. We also cleans the coarse-grained data from the self-edges, assigning $A_{ii}=0$.

\begin{figure}[ht]
\centerline{\includegraphics[width=16cm]{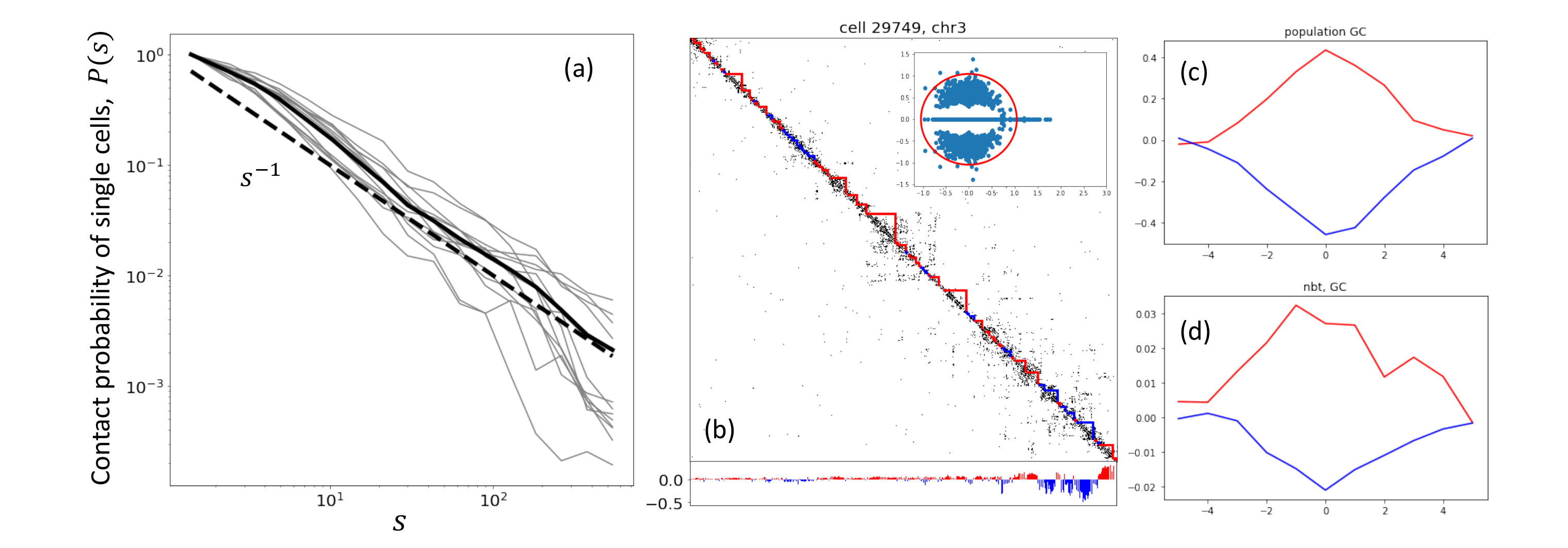}}
\caption{(a) The average contact probability $P(s)$ of single cells (gray) and of the merged cell (solid, black) computed for logarithmically spaced bins with the logfactor 1.4; the fractal globule scaling $P(s)\sim s^{-1}$ is also shown by dashed line for comparison. (b) Annotation of active (red) and inactive (blue) compartmental domains for one of the contact maps (cell 29749, chromosome 3, length $N=492$, 200kb resolution) by the polymer non-backtracking flow operator. Below the map the compartmental signal from the corresponding leading eigenvector of the polymer non-backtracking flow matrix is shown. Inset: the full spectrum of the polymer flow for the same contact map. (c, d) Averaged profiles of the GC content (z-scores) plotted around the centers of the compartmental domains (active - red, inactive - blue) for the population of cells and for a pool of single cells.}
\label{fig:04}
\end{figure}

To determine the background model for our analyses we calculate the contact probability $P(s) = \frac{1}{N-s} \sum_{i=1}^{N-s} A_{i,i+s}$ for each individual single cell and for the merged cell (summing single cells matrices), see \fig{fig:04}a. Resulting dependence turns out to be fairly close to the fractal globule contact probability, $P(s)\sim s^{-\alpha}$ with $\alpha \approx 1$ at scales from $\approx$ 1-2Mb to the end of the chromosome. A shoulder at lower scales around 1 Mb reflects enhancement of the contact probability due to the compartmentalization. Importantly, the fractal globule scaling at the megabase scale is universal across different species and cell types; it is evident in the population-averaged contact matrices in mouse oocytes \cite{flyamer17}, human lymphoblastoid cells \cite{lieberman09} and \textit{Drosophila} cells \cite{ulianov16}. As it was shown in previous Section, in order to extract compartmentalization profile overlaying a specific long-ranged folding, it is crucial to incorporate the respective background contact probability into the polymer model of the stochastic blocks.

Having the background model determined, we construct the polymer non-backtracking flow operator with the variable parameter $\gamma$ and run the iterative clustering procedure to derive the optimal value $\gamma_0$. Similarly to the analyses on the benchmarks, see \fig{fig:03}b, a swift convergence to the optimal value is observed here. The spectrum of the polymer flow operator for the cell 29749, chromosome 3 at $\gamma_0 \approx 0.9$ is shown in the inset of the \fig{fig:04}b. Nineteen isolated eigenvalues on the real axis are separated from the bulk spectrum. As we have shown in the previous Section, this is a quite typical scenario for sparse polymer stochastic block models. In the sparse limit of the polymer SBM, the number of isolated eigenvalues could be much larger than the number of compartments.

%The partition of single cells networks in two compartments has been performed in the leading eigenvector approximation of the different operators. The boundaries of active and inactive domains are determined in accordance with the sign of the respective compartmental signals (see \fig{fig:04}(b) and Fig. S1).

%Interestingly, the operators without polymeric background resolve unexpectedly large clusters, while polymer non-backtracking and polymer modularity detect significantly smaller compartmental domains. Furthermore, the partitions of the polymeric operators are visibly much more adequate to apparent clustering of contacts in a particular cell.

%Annotation of the corresponding bulk Hi-C data is shown in blue for comparison (in accordance with the leading eigenvector of the neutralized observed/expected population-averaged matrix). Most of visible domains correspond to the same A type, while small regions of type B are sandwiched in between the boundaries of A. It is clearly seen that the annotation of the non-backtracking is more relevant to clustering in this particular single cell. At the same time, other annotation (in blue) reflects some average compartmentalization from the ensemble of millions of cells. Such a difference manifests variability of compartments in single cells that has been reported previously \cite{nagano12,flyamer17}.

The partition of the single cells in two compartments has been performed in the leading eigenvector approximation of the different operators. The boundaries of active and inactive domains are determined according to the sign of the respective compartmental signal (see \fig{fig:04}(b) and Supplementary Fig. S1 online). It is known that the gene density is higher in the actively transcribed $A$ compartment, thus, the fraction of GC letters in bins of active compartmental domains needs to be larger than in inactive domains. To validate that the clusters found in single cells respond to the transcriptional domains and are biologically significant, we calculate the GC content profiles around the centers of all $A$ and $B$ domains separately and then take the average of these profiles in each group. The types of the domains were phased in accordance with the leading eigenvector of the bulk data (population Hi-C on embryonic stem cells was used \cite{bonev}; the eigenvector was computed on the observed-over-expected map). We also plot analogous profiles for the leading eigenvector of the bulk data. In absence of direct annotation methods for single cells due to their sparsity, these two measures have been of use to approximate positions of the compartmental domains in single cell Hi-C data \cite{flyamer17}.

As expected, the GC content for the population-averaged map and the bulk E1 vector both have pronounced peaks at the center of $A$ domains and symmetrical dips at the center of $B$ domains with the $z$-score amplitude equal to $0.4$ (GC) and $0.7$ (E1), correspondingly. Single cells profiles demonstrate notably lower amplitudes (see \fig{fig:04}(c,d) and Supplementary Figs. S2, S3 online). However, only the polymer non-backtracking flow yields the annotation with the similar shape and span. Both profiles (for $A$ and for $B$) of the polymer non-backtracking flow fall symmetrically to zero at the same genomic distance, around $4-5$ bins from the centers of domains, which also strikingly coincides with the span of the bulk profiles. This is also compliment to the similarity of the characteristic sizes of compartmental domains determined by the non-backtracking flow operator ($\langle l\rangle \approx$ 2.2 Mb) and domains from the bulk data ($\langle l\rangle \approx$ 1.7 Mb). To test the effect of different $\alpha$, we additionally run the polymer non-backtracking for $\alpha=3/2$, which is the scaling exponent of the contact probability for the ideal chain packing. Comparison of the two values of the parameter is demonstrated in Supplementary Fig. S4 online: the profiles with $\alpha=3/2$ show significantly worse correlation with both GC content and the E1 bulk vector. This is consistent with the slope $\alpha \approx 1$ of $P(s)$ for the set of single cells, \fig{fig:04}a, underscoring the importance of neutralization on the appropriate average polymeric scaling before the clustering.

Note that the partitions of the polymeric operators (non-backtracking, modularity) are visibly much more adequate to apparent clustering of contacts in a particular cell (Supplementary Fig. S1 online). Despite the similarity in compartmental signals from the polymer modularity and from the polymer non-backtracking flow, the sizes of modularity domains are almost twice larger ($\langle l\rangle \approx$ 4.1 Mb) and show negative z-scores of GC content both for the active and inactive compartments. The profile of the E1 vector plotted for the polymer modularity has a similar bell shape, however, it levels at $\approx -0.07$ and stays negative throughout the whole range of the compartmental interval. This is a consequence of sparsity, which results in a limited performance of all traditional spectral methods.

\section{Conclusion}

In this paper we have developed theoretical grounds for spectral community detection in sparse polymer networks. On the basis of suggested polymeric extension of the stochastic block model, we have proposed the polymer non-backtracking flow operator and have proven that its leading eigenvector performs partitioning of a polymeric network into two clusters according the maximum entropy principle. The established connection with the modularity functional provides a computationally efficient tool for the network partitioning and search for the optimal resolution parameter of the partition in polymer networks, which, however, is inferior to the non-backtracking in efficiency for sparse networks.

The proposed theoretical framework is verified by extensive numerical simulations of polymer benchmarks, constructed in order to emulate compartmentalization in sparse chromatin networks. Comparative analyses of different operators on the benchmark has suggested that the polymer flow detects the communities up to the theoretical detectability limit, while all other operators fail above it. At the same time, the amount of isolated eigenvalues of the polymer flow operator can be larger than amount of true communities present in the network, due to frozen linear connectivity that forces the chain to form "blobs" along the chain contour. This result distinguishes the polymer system with thespect the canonical stochastic block model, where the number of isolated eigenvalues of the non-backtracking exactly matches the number of communities.

Analyses of the single cell Hi-C data of mouse oocytes suggests that the non-backtracking walks efficiently split experimental sparse networks into biologically significant communities, characterized by enrichment and depression of the genes density. The sizes of the compartmental domains are fairly close to the sizes of the population-averaged domains. Comparison with characteristics of the domains, inferred by other operators, underscores superiority of the non-backtracking walks in partitioning sparse polymer networks.

In this study we have exploited for the polymer network analysis only the simplest spectral characteristics. More involved ones, e.g. spectral correlators and the level spacing distribution, carry additional information about the propagation of excitations in network. The spectral statistics and non-ergodicity have been discussed in clustered networks in \cite{avetisov, gorsky}. In the context of the gene interactions the spectral statistics has been discussed in \cite{statistics} for the matrices with the real spectrum. The non-backtracking matrices enjoy complex spectrum hence the special means are required to analyze the level spacing in this case. The corresponding tool has been invented recently \cite{statistics2, statistics3}, therefore, the spectral statistics of the polymer non-backtracking flow operator certainly deserves a separate study.

\begin{acknowledgments}
We are grateful to Leonid Mirny, Mikhail Gelfand, Mikhail Tamm, Maxim Imakaev and Nezar Abdennur for valuable discussions on the subject of the paper. KP, AG and SN acknowledge supports of the Foundation for the Support of Theoretical Physics and Mathematics ``BASIS'' (respectively 17-1-2-27-8 for KP, 17-11-122-1 for AG and 19-1-1-48-1 for SN). This work was supported by grants RFBR 18-29-13013 (KP, AG, SN) and RSF 19-14-00016 (SVR, SU). The authors are grateful to CNRS for the organizational and financial support of the publication.

\end{acknowledgments}

\subsection*{Contributions}
K.P. designed the study, performed the benchmark and single cell analyses and wrote the manuscript. A.G., S.N., S.V.R. and S.U. supervised the work and participated in writing.

\subsection*{Competing interests}
The authors declare no competing financial interests.

\section*{Methods. Quadratic form of the polymer non-backtracking operator}

Let us consider a quadratic form involving the operator over the Potts spin variables $g_i, i = 1, 2, ..., N$ and introduce the $2m$-dimensional ($2m$ is the number of edges in the network) vector $u$, such as $u_{i \to j} = g_j$. Then,
\be
R = u^T \mathbf{R}\; u = \sum_{\substack{(i, j) \in E \\ (k, l) \in E}} \mathbf{R}_{i \to j, k \to l}\; g_{j} g_{l}
\label{qflow}
\ee
It can be shown that \eq{qflow} coincides with the quadratic form of the generalized modularity. Let us consider the terms separately. The quadratic form of the first, non-backtracking term, yields
\be
\sum_{\substack{i \to j \\ k \to l}} \frac{\delta_{il}\left(1 - \delta_{jk}\right)}{d_i - 1}\; g_j g_l = \\
\sum_{i, j} \frac{g_i g_j}{d_i - 1} A_{ij} \sum_{k} \left(1-\delta_{0A_{ik}}\right) \left(1 - \delta_{jk}\right) = \\
\sum_{ij} A_{ij}\; g_i g_j
\label{flowterm}
\ee
where the sum over $k$ enumerates the edges of the node $i$ except of the edge $(i, j)$ and, thus, equals $d_i - 1$. Expanding the quadratic form of the second term similarly, we get
\be
\gamma \sum_{\substack{i \to j \\ k \to l}} \left(d_j d_l\right)^{-1}\; P_{jl} \; g_j\; g_l = \\
\gamma \sum_{jl} \left(d_j d_l\right)^{-1} P_{jl} \; g_j\; g_l \sum_{ik} \left(1-\delta_{0A_{ij}}\right)\left(1-\delta_{0A_{kl}}\right) = \\
\gamma \sum_{jl} P_{jl}\; g_{j}\; g_{l}
\label{secterm}
\ee
Collecting \eq{flowterm} and \eq{secterm} together one arrives at
\be
R = u^T R\; u = \sum_{ij} \left(A_{ij} - \gamma P_{ij}\right) g_i\; g_j; \quad P_{ij} = \frac{1}{|i-j|^\alpha}
\label{r}
\ee
which is the quadratic form of the generalized modularity functional, proportional to the entropy of the polymer SBM.

\end{document}